# The Biological Data Sustainability Paradox

Terence R. Johnson & Philip E. Bourne
School of Data Science, The University of Virginia

## Introduction

Biological data in digital form has become *a*, if not *the*, driving force behind innovations in biology, medicine, and the environment. No study and no model would be complete without access to digital data (including text) collected by others and available in public repositories. With this ascent in the fundamental importance of data for reproducible scientific progress has come a troubling paradox.

Diverse data are coming at such a rate, and across many scales (molecules to populations) that with a culture of free availability the system is unsustainable unless there is a greater and more consistent financial investment. Thus far the majority of funds used to build and sustain the biological data ecosystem has come from government agencies. These funders are reluctant to increase spending on the infrastructure needed for data sustainability; this is more than hardware and software, but importantly also includes people. This is particularly true when their missions focus on stimulating innovation in biological research and advancing human health. The data are needed to innovate and to make medical advances - and this costs money, as does new innovative research, which generates yet more data, and so forth. What is the nature of the paradox associated with this circumstance?

## The Paradox

The only simple way out of this situation is to reduce the amount of data being generated (supply) or to spend more money for the curation, preservation and reuse of data (demand). Neither funders as suppliers nor scientists as consumers want to curb progress and neither group wants to spend more money on infrastructure to the detriment of new research. The only long-term answer requires a radical shift towards greater efficiency, public-private partnerships (PPPs) and new economic models, described here as Biological Data Sustainability version 2.0 (BDS 2.0).

Many might immediately disregard BDS 2.0 as absurd, in keeping with the definition of a paradox, but we posit BDS 2.0 as a viable long-term solution. Certainly not the only such solution, but rather intended to provoke thinking towards further alternative models. The current dogma surrounding biological data reflects the fundamental belief of many/most data users that it is a right to have free and unbridled access to high-quality biological data without considering the associated cost, while the data providers see it as part of their mission to be generating the data. Not surprisingly, with these attitudes has come a plethora of data management policies and plans from many funding agencies. What are the options for keeping the current status quo, which we refer to as Biological Data Sustainability version 1.0 (BDS 1.0)?

## BDS 1.0

Under BDS 1.0, a funding agency would need to spend a larger fraction of their budget on sustaining data. As stated, this would seem unpalatable to most users and indeed funders. Why? Most leaders of funding organizations have careers deeply rooted in experimental biology where the costs are tied to reagents, expensive equipment and the support of talented professionals. Data, on the other hand, are perceived as similar to electricity, a utility, by most, supported by a hidden cost paid by a higher authority, for example, university management and available at a flick of a switch. Data are not in their blood as a science unto itself and hence not considered the lifeblood of their work. That said, this culture will change as more of biological discovery is driven by digital data alone, but that transition could take considerable time. Perhaps not at the rate of one funeral at a time as suggested by Max Planck, but slow nevertheless. The emergence of AI as an ever more powerful research tool, which requires accurate and quality data for training might make a mockery of the rate of transition.

If more money is not spent by funders on data, can we be more efficient in all aspects of the data lifecycle - from acquisition to dissemination and reuse? The answer is a definite yes, but there are significant barriers to change, both cultural and political. Stating it bluntly, data are power - whether at the level of the individual research group, institution, or country. Providing data that others want is a form of power that can be to the benefit of the provider. Once the data spigot is turned on, and researchers are drinking, there is a reluctance on both the supply and demand sides to give that up, however inefficient the BDS 1.0 model might be.

How does such inefficiency arise? What starts as a personal collection of data from one or more scientists or a speciality project-borne database,  becomes valuable to others in part because of the affection, skill and knowledge imparted and associated with that data. The sharing of that data, either voluntarily or through data sharing requirements from funders, creates a data resource for which funding is sought and, in the past, often received. Data management experts are hired, features are added and on it goes. The end result is a global collection of disparate and incompatible data sources maintained by individual scientific groups in different countries yet accessed globally, subject to regional restrictions associated with various data types, notably human-derived data. At this point the funders themselves, the researchers maintaining the data resources, the institutions where those resources reside and the countries hosting those institutions are all vested in the 'ownership' of that data resource. Changing that situation would require a change in culture and politics.

The limited successes of BDS 1.0 have come when there is aggregation of disparate data resources under a single organizational structure. Separate data resources are administered by a single body, which leads to exchange of best practices and economies in hardware, software and personnel as well as the provision of a single data governance structure. The European Bioinformatics Institute (EBI) as part of the European Molecular Biology Laboratory (EMBL), the National Center for Biotechnology Information (NCBI) as part of the National Library of Medicine (NLM), in turn part of the National Institutes of Health (NIH),  The Swiss Institute of Bioinformatics, the Center for Information Biology and DNA Data Bank of Japan (CIB-DDBJ) and the Beijing Institute for Genomics (BIG) are examples. Aggregation of data resources mostly respects sovereignty [1] and, as such, there is little cooperation across national boundaries. The same can be said for individual funding agencies within a region [2].

## Enter the Global Biodata Coalition

Recognizing that sustainability of biological data is a global problem, a group of funders have come together to establish the Global Biodata Coalition (GBC) [3]. The GBC is currently legally operated as part of the Human Frontiers Science Program (HFSP) [4], but governed by an independent board of funders. In the meantime, the GBC is seeking independent non-profit status. The GBC represents a noble effort in that it acknowledges the problem (note that PEB is a member of GBC's Scientific Advisory Committee) but is constrained — at the time of writing — by an insufficient number of participating funders around the world. The lack of holistic and truly global funder support to date likely reflects the need to show success to get other funders to join— a Catch-22. Global politics beyond science undoubtedly also leads to GBC's current limited membership.

To broaden its impact, the GBC has been active in evaluating biological data resources, designating a subset as Global Core Biodata Resources (GCBRs). Specifically, open global calls to biodata resources - defined as any biological, life science, or biomedical database that archives research data generated by scientists, or functions as a knowledgebase by adding value to scientific data by aggregation, processing, and expert curation - were made to collect information that would allow the merits of each resource in terms of management and use to be assessed by an independent group of scientists. This builds upon a similar approach that was conducted by ELIXIR in examining European biological data resources [5]. The primary purposes of conferring GBCR designation are: (1) to focus funders' attention on the most critical of the global set of biological data resources in an effort to better understand their parameters for their long-term sustainability; and (2) to stimulate interactions among the community of GBCRs in an effort to have those resources work with global funders in identifying options for their long-term sustainability.  Beyond that, and to solicit broader input, a consultation paper on data sustainability has been made available by GBC as a means to solicit additional input [6]. This paper could be considered an open response to that call.

To summarize thus far. Some global actions are being taken to identify ways to make the data ecosystem more sustainable. This consists of essentially determining what is important and what is not. Such a process has its dangers. A data resource early in its life cycle may be viewed by its funders as serving a small group of users and thus may struggle to attain long-term funding and have their funding cut off. These are difficult calls. Meanwhile, the number of resource closures, or mergers with other resources, has been limited which shows the reluctance of funders to act in the face of opposition by researchers who *must* have access to these data. Researchers are not necessarily concerned that data will become inaccessible, since the storage cost is minimal for all but very large data sets, rather the concern is that additional data perceived as valuable will not be curated and added to the resource. If designating a small group of GCBRs does not help solve this problem, what will? Looking at the private sector and alternative funding models offers clues on a path to BDS 2.0.

# Towards BDS 2.0

A former chief data officer of Airbnb was aghast when asked what data they kept and what data they threw away. "We don't throw away any data, we just figure out how to monetize it," was his response. Taking Airbnb as an example, as we have done in the past for describing service platforms [7], consider the company more from a data sustainability perspective. Airbnb's business is driven by data - more data, means more profit if they use the data effectively. Funding agencies face a different reality. More data means higher costs with no easily measurable advantage. Can the data be monetized? A recent profound scientific development might lead to an answer. *Science* designated AlphaFold from DeepMind as the scientific breakthrough of the year in 2021, because it significantly improved the ability to predict the three-dimensional structure of a protein from its one-dimensional sequence [8]. DeepMind achieved this feat through the use of public data, specifically by data scientists working together with domain scientists in larger groups than are typical in research labs - all with a deep appreciation of engineering and with access to Google compute resources. The potential impact of such an approach on biology and beyond is huge. The verticals of health, agriculture, and energy are obvious examples, each with endless opportunities for monetization and societal benefit.

## Public-Private Partnerships (PPPs)

The obvious monetization model involves the private sector paying for access to public data. This may be short-sighted and likely not to work as previous examples would suggest [9]. On the other hand, paying for commercial cloud resources is already accepted. Paying for the data as well is just the next step. The argument from the private sector is that they pay enough in taxes already. It also does not recognize the benefits of a partnership. In the era of data science, it is naive to think that the major breakthroughs are coming from academia alone. The reality is that they are coming from both academia and the private sector- witness what DeepMind has contributed. Both could be bigger than the sum of the two parts, with society benefiting along the way. If constructed appropriately, data resources could be beneficiaries of such a partnership since both academia and the private sector need unfettered access to the data. Limited examples exist, for example, Open Targets [10], nevertheless, a significant cultural shift is needed to see PPP's as desirable. Consider that shift from the perspective of the different stakeholders, funders, academic research institutions, the private sector and the research scientists in both academia and the private sector.

Government funders are typically constrained in participating in PPPs, although they do have options to participate through affiliated non-profit foundations, (e.g., the Foundation for the National Institutes of Health) that are the entities participating in the PPP. At this time, such partnerships are limited, and funding agencies often fear having their government funding reduced by forming such partnerships. However that is starting to change, witness the NSF's recently formed Technology, Innovation and Partnerships (TIPS) Directorate. Meanwhile, academic research institutions have mixed relationships with the private sector although this varies depending on whether they are public (supported by federal, state or regional governments) or private with little or no direct government funding. For example, public universities are fearful of losing state funding in the face of private sector engagement, although perhaps less so recently with state funding in decline (at least in the US). There is also the ivory tower syndrome that proclaims that association with for-profit entities can hinder academic freedom. The reality is both public and private academic research institutions are businesses, and they need a viable business model to survive. This can drive them to PPP's and will likely do so further in years to come.

Leaders of public data resources are often reluctant, have a sense of entitlement and/or lack the experience and network to engage with the private sector. Some of that reluctance comes from a lingering fear that if they do achieve a PPP then their government funding will be reduced- rather than being rewarded for augmenting their government support with additional private sector resources. Unambiguous support and encouragement from funders could help to alleviate this situation. If currently the private sector will not pay to support data resources and partnerships are slow to form, what are the remaining options?

One option, based on economic theory, is to consider data as a commodity in a marketplace of consumers (researchers), suppliers (data resource operators) and brokers (funders and some consortium, possibly the GBC).

## A Data Economy

It is helpful to reiterate at this point that financial sustainability is not the only criteria by which an intervention or redesign of the market for data should be evaluated. The growth of data in time creates monetary costs of storage, but also of curation, integration, and processing. Any attempt to render the system solvent in a strict financial sense might indeed miss opportunities or even be counterproductive along these other dimensions. A redesign of the market that achieves budget balance but results in vast troves of unusable data will fail to realize a vision not just of free data, but the kind of useful and well-maintained data that enables scientific progress like that exhibited by AlphaFold.

In pursuit of these goals, we propose a data economy that uses cash and an internal currency of data credits to balance the growth of new data deposits with the processing and standardization required to integrate those data with existing and future resources. The idea of a cap-and-trade market in which participants exchange tradable permits goes back at least to Crocker [11], and Kocherlakota [12] first explores how fiat money can be used as literal promises of future value to impose intertemporal trade-offs that improve efficiency. In a related literature in computer science connected to the field of mechanism design [13], the idea of "token economies" to ration the use of scarce resources has led to a large literature considering a variety of rationing and scheduling problems [14], Zahedi et al. [15], Gorokh et al. [16], Guevara et al. [17], Xu and Van Der Schaar[18]. Unlike the premises of these papers, data are not merely an undesirable byproduct of other valuable activities like carbon emissions, and our goal with biological data systems is not to impose a hard limit on the accumulation of data, but rather to ensure balanced and sustainable growth of resources that are labeled, standardized, and integrated into a cohesive whole that enables transformative research.

There are five basic components of the proposed redesign, which is comprised of two markets, for services and credits, and the associated infrastructure (Fig 1):

1. When a party registers with the broker, assuming the GBC, they receive a certain number of credits for free. Every month, a proportion of those credits are automatically replenished, up to the original amount. Thus everyone has access to the system.
2. Whenever a party deposits or downloads data, they must expend credits in proportion to some criteria, possibly the size of the data. Those credits are transferred from the party making the download to the party who deposited the data.
3. In the market for services, parties can solicit data services in return for credits. The buyer posts a request of the form, "For data asset W, I require services X, for which I offer to pay Y data credits, by date Z." The work is completed on a version of data asset W and returned to consortium experts, who then review the work for quality and completeness. If the work is judged acceptable, it is incorporated into the archive, and otherwise rejected.
4. In the market for credits, parties can buy and sell data credits for cash. Buyers post their desired quantity and a maximum price (e.g. "I'd like to purchase 100 credits at any price below $1.30/credit, until 1/1/2024.") while sellers post a quantity and the price they are asking per credit (e.g. "I'm willing to sell 200 credits at a price of $1.00/credit or more, until 1/1/2024."). For any such transaction, the consortium receives a percentage of the asking price in cash as an intermediation "Credit Exchange Fee," for example, 2%. Trades are executed at the seller's asking price plus the exchange fee, as long as that remains below the buyer's stated maximum. So, for the examples above, the buyer would receive 100 credits at a price of $120.00, the seller would receive $100.00 and relinquish 100 credits, and GBC would receive $20.00.
5. The GBC conducts open market operations in the market for credits, making new issues of credits or buying them up to maintain price stability.

The mechanics of the system are designed to create a data economy for which the currency is the data credit, which can be exchanged for cash, data deposits, data downloads, and data improvement services. This is designed so that parties who make popular contributions receive data credits at higher rates, expanding their ability to make contributions and direct resources towards work on data improvement. Any party, however, can provide data improvement services in return for credits, and data depositors and researchers are likely to be the same people at different times. Outside of the data cycle, there is a credit cycle where anyone can participate in the market for credits and then transfer those credits to researchers or depositors. The GBC receives a proportion of the value exchanged in the market for credits, and controls the flow of data credits by issuing new credits to accelerate the system or buying them up to slow it down.

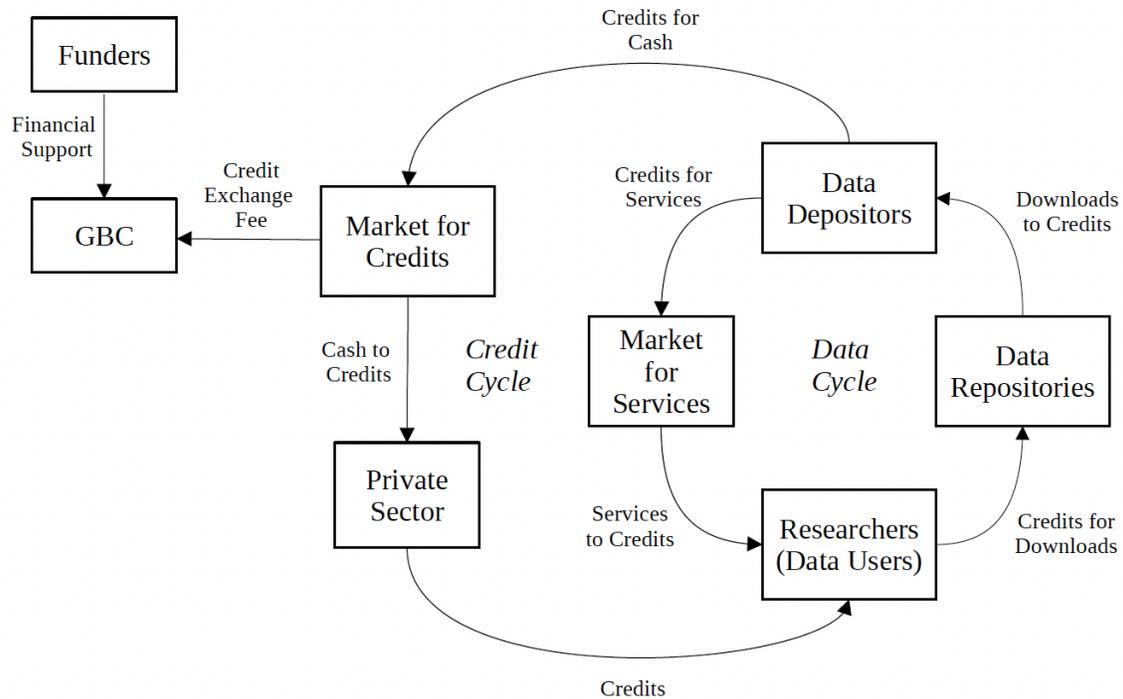

Fig 1. A Data and Service Based Credits Economy

The cornerstone of the system is that the parties depositing popular and valuable data are rewarded with credits every time a download occurs: This is a "depositor dividend" for providing widely used and valued data. These will likely be academic units or elite labs whose work is pushing forward various boundaries across disciplines. Returning economic power to them through the accumulation of data credits has two consequences: They are the most likely to receive monetary payments from parties who need to purchase credits in the market for credits, and they have substantial resources to pursue further data curation and standardization of their data or related data in the market for services. As a result, the problem of selecting what data to curate and improve is decentralized, with the parties who provide the most useful contributions to the archive having significant influence over curation and integration. Conversely, agents are dis-incented from depositing low quality data that has little potential: Depositing costs data credits and is unlikely to yield dividends in the future because few people will download these data. This reduces the burden on the consortium to curate and evaluate the quality of data: The system is designed to align the interests of the depositors with the consortium to incent data deposits that are likely to be impactful, and dis-incent depositing data without considering the benefits it might yield to others, or merely to list the activity on a CV. By tuning the intermediation fee up and down, the consortium raises more or less money, and that money comes disproportionately from labs and agents in the space who are likely to have deep pockets (i.e. be non-academic and profit-maximizing).

Depositing data and obtaining data services, including downloading, requires expending credits. If credits naturally flow to the parties providing the best data, then credits naturally flow from the parties who are the most data poor. For private sector actors whose work is largely proprietary and cannot be deposited and shared, they will likely pay into the system with cash. For academic actors whose work is public but does not result in an easy flow of credits from downloads, they can participate in the market for services. This is ideal for a variety of reasons. First, data labeling and wrangling requires expert knowledge but is, in general, time-consuming, tedious, and quickly leads to burnout. Doing short stints of work on consortium projects to generate data credits becomes a part of contributing to data quality at large and can increase activity at under-utilized labs. Otherwise harnessing expertise to do these kinds of tasks would be difficult or impossible. Indeed, this kind of work is ideal for labs who are getting started with new data resources related to a given cleaning task: They can get access to the data, contribute to improving its quality, and earn data credits, all at the same time, despite being potentially financially and materially disadvantaged at their institution. Second, the ideal people to be doing expert tasks like labeling and standardizing are the people who work with the data and understand its use on a deep level. By tying access to data depositing and downloading to contributing to data quality, various actors and labs become partners in creating higher quality, more integrated, and better curated data for everyone. Third, reputation becomes important: Labs that get a reputation for failing to deliver high quality data work upon consortium expert review can easily be punished through the system by restricting their access to data services or downloads, thus ensuring that the work done to improve the data is truly improving it and not just churning the data credit cycle. All of these features together create a robust market for data services.

The market for credits plays a crucial role in the system overall. First, it expedites the acquisition or deposition of data for parties that have to act quickly and have sufficient resources. Second, it monetizes the value of data indirectly by establishing an exchange rate between money and data credits, and what data credits provide: downloading, depositing, and cleaning. Third, it compensates actors for depositing high quality data, enabling future work. Fourth, it provides a means to fund the system overall by indirectly taxing the exact parties who are not making data contributions themselves but reaping commercial awards from their use.

Managing the market for credits requires maintaining financial stability. First, data credits are constantly being replenished globally according to #1, up to a certain level. This means that there is some natural inflation in the system that allows access to everyone, but rate limits their downloading. To offset the inflation, some credits are retired whenever a download occurs. That makes it possible to achieve, at least in principle, a steady-state price level. Second, the consortium must act as a kind of regulatory body, similar to the Federal Reserve or other monetary authorities that act to stabilize a currency, ready to purchase or withdraw credits from the system if the price is falling to levels that are unsustainably low, or release credits if prices are becoming too high. The goal is an orderly and predictable flow of value from all the parties involved, with a minimum of uncertainty or disruption over time, which would undermine the creation of the system as a method of supporting scientific research. In general, having too few credits will lead to high prices for data credits and slow the pace of research, while having too many credits will lead to low prices but a financially unsustainable system. A sustainable, reliable, steady state system requires balancing automatic deposits with credit retirement.

There are then two paths to acquiring data credits: Purchasing them at market rates or acquiring them by doing data work that improves the value of the global archive.

The apparent trade-off in rate-limiting downloading by requiring data credits would appear to tax progress. This ignores that the main constraint on scientific advancement is actually data quality. As the stock of raw data grows rapidly in relation to the data that are curated, labeled, and integrated, the rate of progress can actually slow down because researchers must devote resources to searching and processing the data for their own purposes. Worse, these improvements might not make it back to the archive at the end or be harmonized with other activities by other labs. The market for data services strikes a balance between centralized data storage and decentralized data improvement. So the relevant tension is not solely a trade-off between download accessibility/progress and financial sustainability, but rather between the unchecked growth of the data and its cultivation as a useful resource. Maintaining totally free downloads and deposits appears to superficially maximize access to a resource, but what good is a vast trove of unusable data files? By intervening to incent deliberate curation and cultivation illustrates how opposition to monetizing the data misses the real challenge, which is not merely conservation of data files but curation of high quality and tightly integrated data resources that enable scientific progress at every level.

Overall, this design provides mechanisms to balance depositing and curating data, commerce and science, data exploitation and exploration, and decentralizes a number of key, difficult tasks from the consortium to the actors in the system. This description of the mechanics raises a number of questions.

How should credits be initially distributed? The simplest answer is to look at historical patterns of data downloads. Reward initial credit allocations in proportion to the share of total historic downloads (or the log share if tails are extremely long, as is likely the case). An alternative or complement is to auction credits to raise one-time funds for the consortium, similar to an Initial Public Offering (IPO) of stock.

Can downloading data be made completely free, removed entirely from the data credit economy? There are legitimate reasons to champion the ideal of free data. There are, however, other values that communities can also endorse. Nothing is truly free, and the value of the data is increasingly the fruits of a complex community collaboration, not the work of a single grant or lone labs. Part of building a sustainable future is cultivating appreciation of these priceless resources by contributing to them. No one should feel unconditional entitlement towards these resources, but rather gratitude for the previous and ongoing efforts of their peers that make progress possible. While there is a monetary, pecuniary shortcut to downloading data immediately, there is a longer but more socially productive path to data resources which simultaneously improves data quality for everyone. Making data downloads free reduces the power of the incentives provided, ultimately undercutting the system and making it less likely to succeed.

Is this sustainable? More specifically, why is this more sustainable than current, government-funded models? It leverages the entire community of scientists, to one degree or another, to create a shared data economy. As activity scales up, so presumably do receipts, and financial support for the consortium. Since data services are integrated into the system, all aspects of the costly and complex problem of data curation and management are addressed through a decentralized economy that directs resources towards the users' greatest utility. Many difficult choices about how to improve, label, curate, and distribute data will be endogenously resolved by how data credits are spent on services. If there are competing standards of data cleaning or governance, they can simply compete for users' data credits.

Is this equitable? Everyone gets access to the data, but they decide on the price they pay. Parties for whom the cheapest route to the data is to pay for it can compensate the depositors indirectly and help fund the consortium through the market for credits. If financial resources are scarce, labs can instead take on data services work instead. This distributes the burden of curating, maintaining, and improving the archive across all of its users. By design, the depositors who enable the most work are rewarded and given influence over the development of the data resources. Similarly, by design, parties that are data poor or cash poor contribute through data services like curation and standardization. This builds an ecosystem of collaborators who are differentially contributing to improving the available data resources depending on their relative strengths and needs.

Why have a market for credits at all? Time is the most precious resource for everyone, and particularly in research. The market for credits is really a way of purchasing time, rather than providing data services or making significant contributions to the community. This is one way of conceptualizing the cultural problem at the bottom of conflicts over openness and freeness of resources: Some groups spent significant time and energy collecting and curating data, and other people want to forego those costs and go straight to the rewards. That can be arranged through this system, either through monetary expenditure or non-pecuniary efforts that simultaneously improve the quality of data available for everyone. The system asks, in return, that people participate in the mutually beneficial work of building an ecosystem of markets and services that incrementally improve the data that everyone depends on.

Some datasets are increasingly the product of work by groups, or even take on a life of their own after deposit and become a composite work of many contributors over time, many of whom might not even know one another. How should that be handled? This is a complex question, but a glib answer is available. Any version of the data is available for some period of time, and its desirability determined by the change in download levels. That change can be decomposed using the Shapley Value to attribute which parts of the overall download levels are attributable to the previous version, which is attributable to the new deposit or data work. If some dataset truly becomes the work of such diverse hands that it can no longer be considered the work of any given agent, it might be placed in the "public domain" of the system and its downloads no longer yield credits to any particular party.

## Conclusion

The concept of a data economy embodied in Biological Data Sustainability 2.0 is a radical departure from how biological data resources are currently sustained. The concept embodies ideas from fields that operate very differently from biology - and that is precisely the point. The current problematic situation would benefit enormously from the engagement of more people from disparate fields thinking about the problem from new perspectives. Until such new thinking gains traction, the solutions of putting ever more money into sustaining biological data resources writ large and possibly stopping the funding of the less favored data resources will persist.

## Acknowledgements

The concept of data credits arose in discussions with Henning Hermjakob. Thanks to Ron Appel, Cristine Durinx, Eric Green, Varsha Khodiyar, Paula Mabee, Nicky Mulder, Brian Nosek, Philippe Sanseau and Yongbaio Xue for constructive input.